\documentclass[11pt, a4paper]{article}

\usepackage[margin=1in]{geometry}
\usepackage{graphicx}
\usepackage{amsmath}
\usepackage{booktabs}
\usepackage{multirow} 
\usepackage[table]{xcolor}
\usepackage{times} 
\usepackage{pbox}
\usepackage{amssymb}  
\usepackage{amsfonts} 
\usepackage{makecell}
\usepackage{caption} 
\usepackage{hyperref} 
\usepackage{authblk}

\title{\textbf{A Hybrid CNN-VSSM model for Multi-View, Multi-Task Mammography Analysis: Robust Diagnosis with Attention-Based Fusion}}

\author[1]{Yalda Zafari} 
\author[1]{Roaa Elalfy}
\author[1]{Mohamed Mabrok}
\author[2]{Somaya Al-Maadeed}
\author[3]{Tamer Khattab}
\author[4, 5]{Essam A. Rashed}

\affil[1]{Department of Mathematics and Statistics, Qatar University, Doha, Qatar}
\affil[2]{Department of Computer Science and Engineering, Qatar University, Doha, Qatar}
\affil[3]{Department of Electrical Engineering, Qatar University, Doha, Qatar}
\affil[4]{Graduate School of Information Science, University of Hyogo,
Kobe 650-0047, Japan}
\affil[5]{Advanced Medical Engineering Research Institute, University of Hyogo, Himeji, 670-0836, Japan}

\date{}

\begin{document}

\maketitle

\begin{abstract}
Early and accurate interpretation of screening mammograms is essential for effective breast cancer detection, yet it remains a complex challenge due to subtle imaging findings and diagnostic ambiguity. Many existing AI approaches fall short by focusing on single-view inputs or single-task outputs, limiting their clinical utility. To address these limitations, we propose a novel multi-view, multi-task hybrid deep learning framework that processes all four standard mammography views and jointly predicts diagnostic labels and BI-RADS scores for each breast. Our architecture integrates a hybrid CNN–VSSM backbone, combining convolutional encoders for rich local feature extraction with Visual State Space Models (VSSMs) to capture global contextual dependencies. To improve robustness and interpretability, we incorporate a gated attention-based fusion module that dynamically weights information across views, effectively handling cases with missing data. 
We conduct extensive experiments across diagnostic tasks of varying complexity, benchmarking our proposed hybrid models against baseline CNN architectures and VSSM models in both single-task and multi-task learning settings. Across all tasks, the hybrid models consistently outperform the baselines. In the binary BI-RADS 1 vs. 5 classification task, the shared hybrid model achieves an AUC of 0.9967 and an F1-score of 0.9830. For the more challenging ternary classification, it attains an F1-score of 0.7790, while in the five-class BI-RADS task, the best F1-score reaches 0.4904. These results highlight the effectiveness of the proposed hybrid framework and underscore both the potential and limitations of multi-task learning  for improving diagnostic performance and enabling clinically meaningful mammography analysis.

\end{abstract}

\newpage

\section{Introduction}

Breast cancer is the most commonly diagnosed cancer and a leading cause of cancer-related deaths among women globally \cite{giaquinto2022breast}. Early detection through mammography screening is central to reducing mortality \cite{fiorica2016breast}. However, interpreting mammograms is challenging due to subtle malignancy indicators (e.g., microcalcifications, distortions) and high inter-observer variability, leading to false positives/negatives and motivating the development of AI-assisted diagnostic tools.

Conventional 2D mammography has inherent limitations in accurately capturing 3D breast anatomy. Standard mammography includes two views per breast, cranio-caudal (CC) and medio-lateral-oblique (MLO), which clinicians analyze using both \textit{Ipsi-lateral} (within-breast) and \textit{Contra-lateral} (between-breast) comparisons. Findings are categorized via the BI-RADS system: 0 (incomplete), 1 (negative), 2 (benign), 3 (probably benign), 4 (suspicious), 5 (highly suggestive of malignancy), and 6 (biopsy-proven malignancy). While biopsy confirms diagnosis, mismatches between BI-RADS scores and biopsy-based diagnosis can arise from interpretive errors or imaging limitations \cite{comstock2020comparison}.

The advent of Deep Learning (DL) has marked a new era in Computer-Aided Diagnosis (CAD) systems \cite{hussain2024revolutionising}. Convolutional Neural Networks (CNNs) dominate due to their strength in learning local spatial features from pixel data \cite{yu2021convolutional}, though their limited receptive fields restrict global context modeling. To address this, vision transformers \cite{dosovitskiy2020image, liu2021swin, shamshad2023transformers, zafari2024transformers} and Vision State Space Models (VSSM) \cite{liu2024vmamba, yue2024medmamba, ruan2024vm} have emerged, offering global feature modeling. Combining CNNs with these architectures enhances representation quality and diagnostic accuracy.

A clinically meaningful CAD system for mammography analysis must handle multi-view inputs (mimicking radiologist workflows), support missing views, and allow continuous improvement through human feedback. Integrating multi-task capabilities (such as BI-RADS assessment and malignancy prediction) closely aligned with clinical practice, offers significant potential for fine-tuning model performance at various diagnostic stages. This multi-task approach enhances the system’s adaptability and utility across different levels of clinical decision-making.

Most deep learning multi-view mammography models emphasize two core tasks: feature extraction and cross-view fusion. CNN-based architectures remain the dominant backbone for feature extraction, although recent studies have also explored the use of vision transformers. A summary of various deep learning-based approaches for multi-view mammography classification is presented in Table \ref{tab:research_summary}. Commonly, models use ipsi-lateral analysis (CC and MLO per breast), with simple fusion techniques like concatenation or averaging \cite{truong2023delving}. The majority of research efforts have focused on single-task classification, primarily distinguishing malignant cases from benign or normal ones. These models were often evaluated and fine-tuned on external unseen datasets, where they maintained comparable performance \cite{isosalo2023independent, yang2023mammodg}. Where biopsy data is unavailable, BI-RADS labels serve as surrogate ground truth, though inconsistencies can affect evaluation \cite{sarker2024mv}.

Recent works have explored cross-attention modules to capture inter-view relationships \cite{van2021multi}, extending to four-view models under side-invariant and side-variant paradigms \cite{manigrasso2025mammography}. While these methods enhance contextual learning, they also introduce significant computational overhead due to quadratic complexity, especially with high-resolution mammography images. This makes early-layer integration difficult and reduces feasibility in real-world clinical environments. Additionally, the inclusion of additional parameters increases model complexity, thereby requiring large-scale datasets that capture a diverse range of features influencing performance \cite{soliman2024deep, dosovitskiy2020image}. Furthermore, such models often assume complete view availability and lack built-in mechanisms for handling missing data. Notably, to the best of our knowledge, existing studies have not proposed strategies or conducted evaluations to address scenarios involving incomplete view data.

\begin{table*}[!t]
\centering
\resizebox{\textwidth}{!}{%
\begin{tabular}{m{1.5cm} m{8cm} m{8cm} m{2cm}}
\toprule
\textbf{Reference} & \textbf{Dataset \{Classes\}} & \textbf{Approach} & \textbf{Results (AUC)} \\
\midrule
\cite{van2021multi} &
CBIS-DDSM (mass subset) \{Benign vs. Malignant\} \cite{lee2017curated} &
Dual-view mammography classification using ResNet-18 for feature extraction and a transformer-based cross-view attention mechanism. &
0.803 \\ \hline

\cite{chen2022transformers} &
Private dataset \{Non-malignant vs. Malignant\} &
Four-view images partitioned into patches processed via local transformers (shared weights), followed by global transformer-based fusion. &
0.814 \\ \hline

\cite{truong2023delving} &
\pbox{8cm}{VinDr-Mammo \{BI-RADS(2) vs. BI-RADS(4,5)\} \cite{nguyen2023vindr} \\ CMMD \{Non-malignant vs. Malignant\} \cite{cai2019breast, wang2016discrimination}} &
Analysis of multiple fusion strategies (average and concatenation) at various ResNet-18 layers for dual-view mammography classification. &
\makecell[l]{0.7535 \\ 0.8416} \\ \hline

\cite{yang2023mammodg} &
\pbox{8cm}{CBIS-DDSM \{Benign vs. Malignant\} \cite{lee2017curated} \\ CMMD \{Benign vs. Malignant\} \cite{cai2019breast, wang2016discrimination}} &
Dual-view model based on ResNet-18 with added channel- and feature-attention modules across various network layers. Features from ipsi-lateral views are further aggregated using a transformer-based global encoder. &
\makecell[l]{0.7798 \\ 0.8181} \\ \hline

\cite{isosalo2023independent} &
Private dataset \{Non-malignant vs. Malignant\} &
Four-view analysis using ResNet-22 for feature extraction and late-stage fusion via concatenation. &
\makecell[l]{R-MLO: 0.82 \\ R-CC: 0.85 \\ L-MLO: 0.84 \\ L-CC: 0.83} \\ \hline

\cite{sarker2024mv} &
\pbox{8cm}{CBIS-DDSM \{Benign vs. Malignant\} \cite{lee2017curated} \\ VinDr-Mammo \{BI-RADS(1–3) vs. BI-RADS(4,5)\} \cite{nguyen2023vindr}} &
A dynamic attention fusion module using shifted windows to integrate information across dual-view mammograms. &
\makecell[l]{0.6643 \\ 0.9608} \\ \hline

\cite{liu2021act, manigrasso2025mammography} &
CSAW case-control subset (Karolinska) \cite{dembrower2020multi}, DDSM \cite{heath1998current}, Synthetic data \{Non-malignant vs. Malignant\} &
An anatomy-aware framework based on graph convolutional networks. It models intra-breast geometric relationships (ipsi-lateral) and contra-lateral similarities, followed by a correspondence reasoning module for improved results. &
0.844 \\ \hline

\cite{manigrasso2025mammography, van2021multi} &
CSAW case-control subset (Karolinska) \cite{dembrower2020multi}, DDSM \cite{heath1998current}, Synthetic data \{Non-malignant vs. Malignant\} &
Four-view analysis using Swin Transformer \cite{liu2021swin} for feature extraction and cross-attention mechanisms to capture inter-view relationships. &
0.767 \\  
\bottomrule
\end{tabular}
}
\caption{Summary of recent deep learning approaches for multi-view mammography classification. Reported AUC values correspond to datasets used for model training or fine-tuning. Results from datasets used solely for evaluation are excluded.}
\label{tab:research_summary}
\end{table*}

Transformers’ limitations in scalability have prompted interest in Mamba, a state space model architecture that captures long-range dependencies with linear complexity \cite{gu2023mamba}. Building on this, we propose a hybrid CNN-VSSM framework for multi-view mammography. The model jointly predicts malignancy and BI-RADS scores, using a pretrained CNN encoder to extract local features followed by VSSM layers for global context modeling. To integrate multi-view inputs effectively and handle missing views, we introduce a gated attention fusion module, enhancing both interpretability and robustness. The primary contributions of this work are therefore five-fold:

\begin{itemize}
    \item We develop a novel multi-view, multi-task, quad-head architecture that uniquely processes the complete four-view mammography study (L-CC, L-MLO, R-CC, R-MLO) to simultaneously and independently predict diagnostic outcomes and BI-RADS assessments for both breasts in a single forward pass.
    \item We introduce a new Hybrid CNN–VSSM architecture, which effectively cascades convolutional neural networks with visual state space models, combining the strengths of CNNs in local feature extraction with the global context modeling capabilities of SSMs.
    \item We incorporate a gated attention-based fusion mechanism that intelligently learns to assign dynamic importance to each view, enabling the model to construct a more robust for missing data and context-aware fused representation.
    \item We perform a comprehensive empirical comparison across a range of models, examining their effectiveness on prediction tasks with varying levels of clinical ambiguity and uncertainty.
    \item We investigate single-task versus multi-task training strategies for simultaneous prediction of diagnostic outcomes and BI-RADS scores, two clinically related yet occasionally divergent tasks, to better understand the benefits and trade-offs of joint learning in this context.
\end{itemize}

\section{Architectures}
Let $\mathcal{X}$ be the space of mammography image sets, where each image set consists of four standard views: R\_CC, R\_MLO, L\_CC, and L\_MLO. Let $\mathcal{Y}_1 = \{0, 1\}$ and $\mathcal{Y}_2 = \{1, 2, 3, 4, 5\}$ denote the label spaces for two classification tasks, such as malignancy classification and BI-RADS assessment, respectively. Each data point is represented as a triplet $(x_i, y_{i, s}^{(1)}, y_{i, s}^{(2)})$, where $x_i = \{x_i^{\text{R\_CC}}, x_i^{\text{R\_MLO}}, x_i^{\text{L\_CC}}, x_i^{\text{L\_MLO}}\} \in \mathcal{X}$ denotes the set of four-view images for patient $i$, $y_{i, s}^{(1)} \in \mathcal{Y}_1$ is the label for the first task for side $s \in {\text{R, L}}$, and $y_{i, s}^{(2)} \in \mathcal{Y}_2$ is the label for the second task. Our goal is to learn a multi-task multi-view classifier $f_{\theta} : \mathcal{X} \rightarrow \mathcal{Y}_1 \times \mathcal{Y}_2$ that achieves high predictive performance on both tasks for both sides.

\subsection{Feature Extraction}

\subsubsection{Baseline: CNN-based}
\begin{figure*}
    \centering
    \includegraphics[width=1\linewidth]{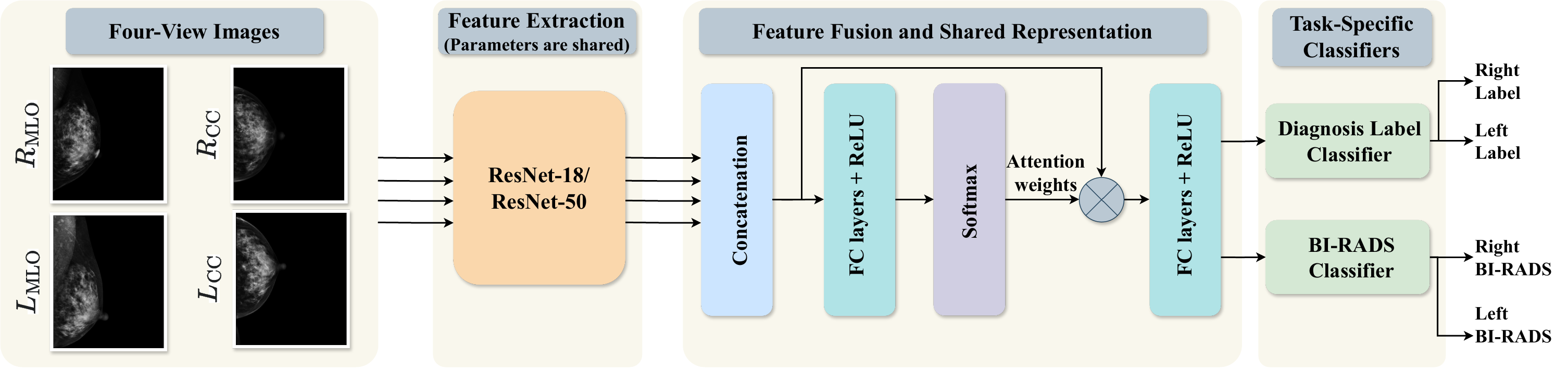}
    \caption{The ResNet-based architecture. A single, shared ResNet-18/ResNet-50 backbone acts as the feature extractor for all four input views. The resulting feature vectors are then intelligently weighted and combined by the gated attention fusion module and then fed into the multi-task classifier heads.}
    \label{fig:resenet}
\end{figure*}

Given their widespread use in medical image classification, especially mammography, we adopted ResNet-18 and ResNet-50 \cite{he2016deep} as baseline CNN architectures. ResNet-18 was selected for its proven effectiveness in prior studies, while ResNet-50 allowed us to examine the impact of increased model complexity. Both networks were initialized with ImageNet-pretrained weights to leverage transfer learning. Consistent with standard multi-view mammography approaches, parameters were shared across views to ensure uniform feature extraction and reduce the number of trainable parameters. This setup provides a strong and conventional benchmark for evaluating our proposed multi-view, multi-task framework. In the multi-task configuration, both classifier heads were active, whereas single-task training used only the relevant output head. Figure \ref{fig:resenet} depicts the overall architecture.

\subsubsection{Vision State Space Model (VSSM)}
State Space Models (SSMs) are powerful tools for modeling dynamic systems and sequences via latent hidden states influenced by inputs \cite{gu2023mamba}. In machine learning, SSMs have recently gained traction as efficient alternatives to recurrent and attention-based models for capturing long-range dependencies \cite{gu2023mamba}. In continuous-time linear SSMs, the system dynamics are governed by:

\begin{align}
h^{'}(t) &= A h(t) + B x(t) \quad \text{(State transition)},  \notag \\
y(t) &= C h(t) + D x(t) \quad \text{(Readout)} 
\end{align}

where \( h(t) \in \mathbb{R}^{N} \) is the hidden state,  \( x(t) \in \mathbb{R}^{d_{\text{in}}} \) is the input vector, and \( y(t) \in \mathbb{R}^{d_{\text{out}}} \) is the output vector. The matrices \( A \in \mathbb{R}^{N \times N} \), \( B \in \mathbb{R}^{N \times d_{\text{in}}} \), \( C \in \mathbb{R}^{d_{\text{out}} \times N} \), and \( D \in \mathbb{R}^{d_{\text{out}} \times d_{\text{in}}} \) correspond to the state transition, input-to-state, state-to-output, and direct input-output (skip connection) mappings, respectively. While foundational, this formulation has been enhanced for deep learning via models such as Structured State Space Sequence Model (S4) \cite{gu2021efficiently}, which efficiently captures long-range dependencies using structured matrices like HiPPO. Rather than updating the hidden state iteratively, S4 applies a convolutional kernel:

\begin{equation}
y_t = \sum_{i=0}^{t} K_{t-i} x_i
\end{equation}

where the convolution kernel is:

\begin{equation}
K_i = C A^i B
\end{equation}

To improve computational efficiency, S4 computes this convolution using the Fourier domain as follows: 

\begin{equation}
y = K * x \quad \text{(efficiently via FFT),}
\end{equation}

where \( K_i \in \mathbb{R}^{d_{\text{out}} \times d_{\text{in}}} \) and the kernel length can be up to the sequence length. This approach leverages the Fourier domain for efficient computation, allowing for linear complexity with respect to the sequence length \( L \). Building on S4, the Mamba model \cite{gu2023mamba} extends the expressiveness of SSMs by introducing data-dependent and dynamic parameters. Mamba enhances the S4 framework, referred to as S6 in this context, by conditioning the state-space matrices on the current input. This adaptive mechanism enables selective memory, allowing the model to dynamically determine what information to retain or discard based on the input at each timestep. In Mamba, the state update is simplified by omitting the $D$ matrix, as skip connection, which can be added back easily. A key component of Mamba’s implementation is the discretization of continuous-time dynamics, making the model compatible with deep learning workflows. This is achieved by introducing a time-scale parameter $\Delta$ and applying a discretization rule to approximate the continuous dynamics:

\begin{align}
  \overline{A} &= exp(\Delta A), \notag  \\
   \overline{B} &= {(\Delta A)}^{-1}(exp(\Delta A) - I)  (\Delta B).
\end{align}
The discrete version takes the following form: 
\begin{align}
    h_t &= \overline{A}h_{t-1} + \overline{B}x_t, \notag  \\
   y_t &= Ch_t.
\end{align}
The convolution operator in the discrete case is given  as follows: 
\begin{align}
    \overline{K} &= (C\overline{B}, C\overline{AB}, ..., C\overline{A}^{L-1}\overline{B} ), \notag  \\
   y_t &= x * \overline{K}
\end{align}

\begin{figure*}
    \centering
    \includegraphics[width=1\linewidth]{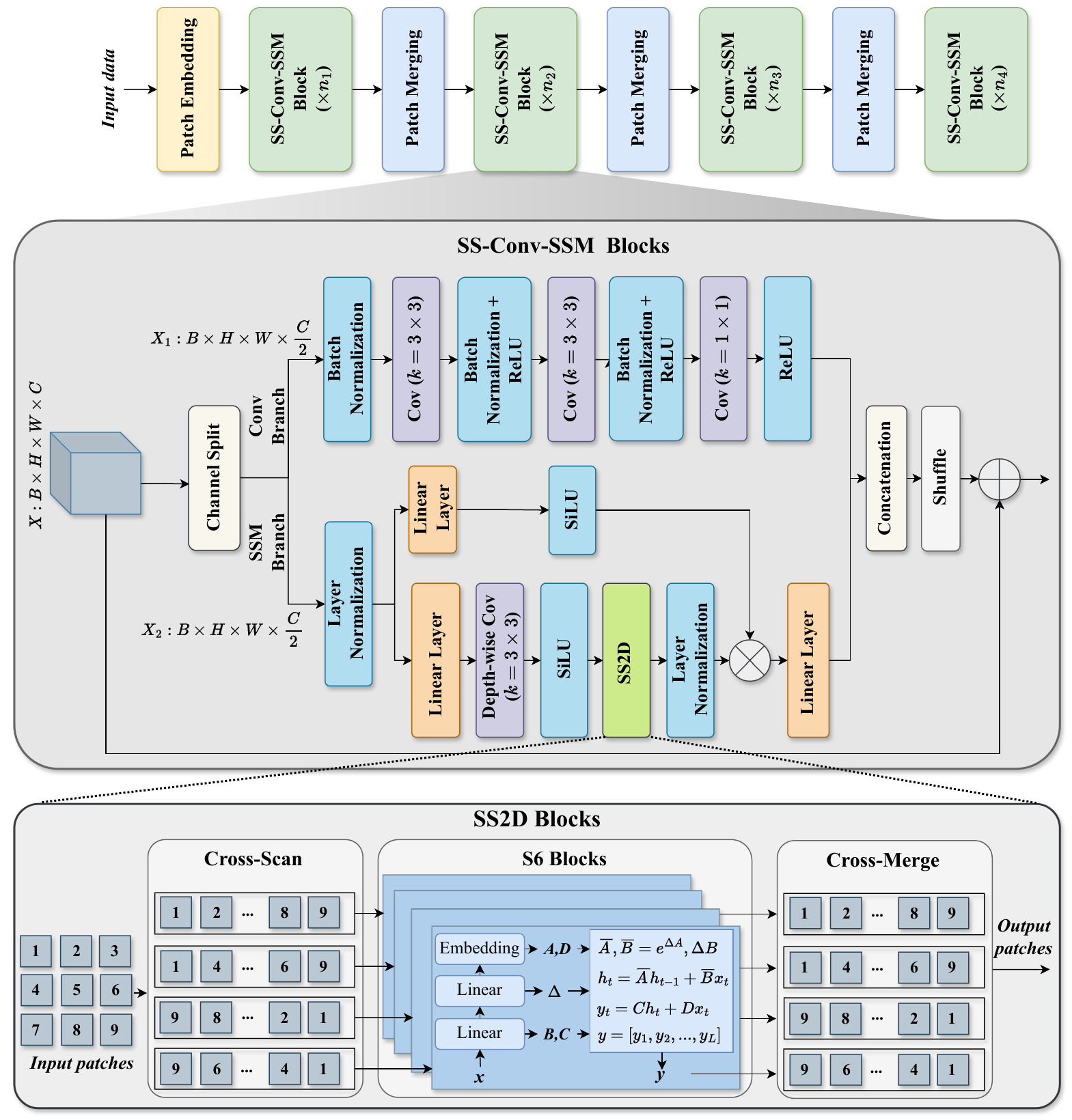}
    \caption{The VSSM-based architecture. The proposed architecture comprises multiple hybrid convolution-SSM blocks, each containing parallel convolutional and SSM-based branches. This design enables the preservation of local spatial details while facilitating the extraction of global contextual information. The SSM modules incorporate 2D Selective Scan (SS2D), which scans the input using diverse schemes in parallel and subsequently merges the results. This approach helps mitigate the loss of temporal information that typically occurs during image sequencing when feeding data into SSM models.}
    \label{fig:vssm}
\end{figure*}

Extending this approach to visual domains, the Vision Mamba (VMamba) framework introduces Visual State Space (VSS) modules tailored for image data. 2D Selective Scan (SS2D) mechanism \cite{liu2024vmamba} is the core idea to this extension, which incorporates four distinct scanning strategies over extracted image patches. These diverse scan paths allow the model to capture rich spatial dependencies often lost when flattening images into 1D sequences. By retaining two-dimensional structure and incorporating directional biases, SS2D enhances spatial modeling capacity. Figure \ref{fig:vssm} provides an overview of the SS2D mechanism and the VSSM architecture.

In this study, we employed the architecture from \cite{yue2024medmamba}, which integrates CNN and VSSM branches to jointly capture local and global features in medical images. The input is split channel-wise: one part passes through convolutional layers, the other through VSSM blocks as defined in \cite{liu2024vmamba}. The outputs from both branches are concatenated to form a unified feature representation (see Figure \ref{fig:vssm}).

\subsubsection{Proposed Hybrid CNN-VSSM}
The proposed architecture in this study is a hybrid CNN-VSSM model that synergistically combines the complementary strengths of CNNs and SSMs. The model follows a cascaded structure comprising two primary components. Initially, the input image is processed through the early layers of a pretrained ResNet-18, which serves as an efficient and robust feature encoder. This stage leverages strong inductive biases and pretrained knowledge of the CNN to convert raw pixel data into mid-level feature maps rich in local patterns, textures, and structural information. 

\begin{figure*}
    \centering
    \includegraphics[width=1\linewidth]{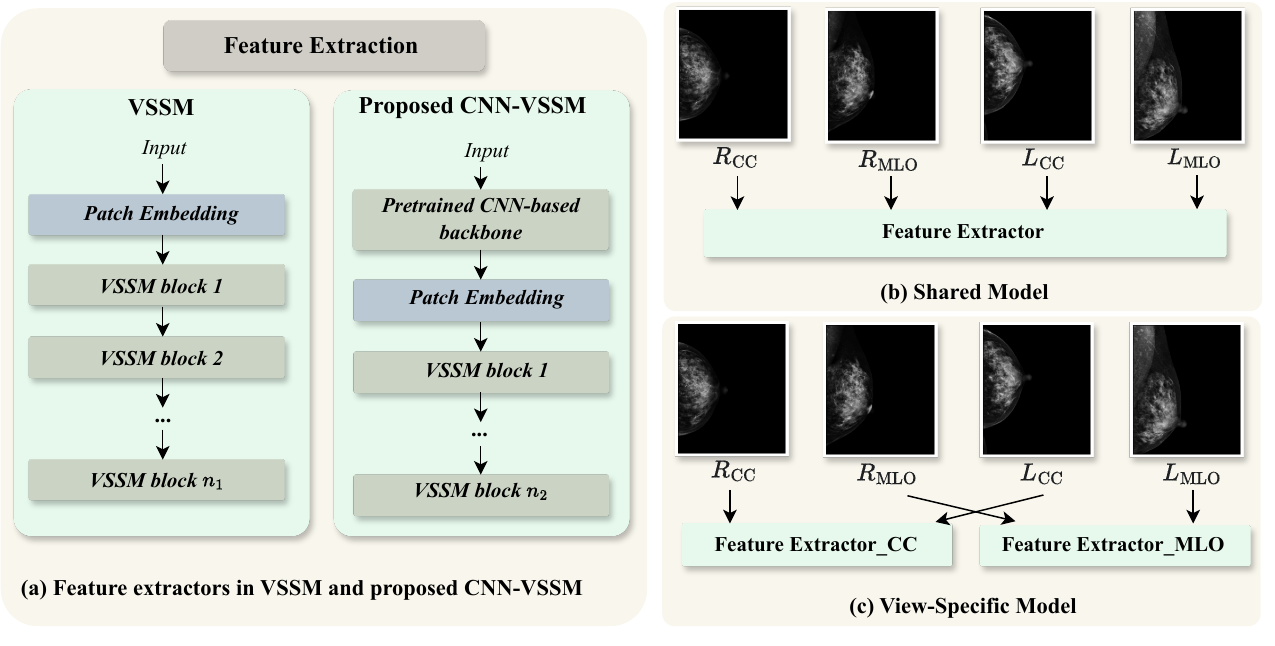}
    \caption{(a) The feature extractor in both the VSSM-based model and the proposed CNN-VSSM architecture. The CNN-VSSM utilizes the initial layers of a pretrained ResNet-18 to serve as a rich encoder for local spatial information. (b) The shared-model input strategy, where all image views are processed through a common encoder. In this setup, model parameters are shared across views to promote generalization. (c) The view-specific model configuration, which consists of two parallel backbone networks with identical architectures, one dedicated to CC view images and the other to MLO views. This design enhances the model's capacity to adapt to view-dependent anatomical variations and global contextual differences.}
    \label{fig:resnet-vssm}
\end{figure*}

The resulting feature map is then treated as a sequence of feature tokens and passed into the subsequent stages of the VSSM. Rather than learning directly from pixels, the VSSM component is responsible for high-level contextual reasoning, capturing long-range dependencies and modeling global relationships while maintaining critical local information embedded in the CNN-extracted features. Figure \ref{fig:resnet-vssm} illustrates the differences between standalone VSSM models and the proposed hybrid CNN-VSSM model in terms of their feature extraction pipelines.

To fully exploit the capabilities of VSSM in modeling global context, we explored two input strategies: (1) a shared model in which the same parameters are used across all views, and (2) a view-specific model architecture consisting of two parallel branches with identical backbones but separate weights. The view-specific approach is motivated by the anatomical and visual differences between views; for instance, the MLO view often includes the pectoralis major muscle, which is not visible in the CC view.

\subsection{Feature Fusion}

A key element in multi-view architectures is the fusion strategy that combines features from individual views into a unified representation. A typical baseline is feature concatenation, which is simple and robust to missing views. However, treating all views equally is suboptimal, as clinical relevance often varies across projections (e.g., a lesion may be most conspicuous on the MLO projection). To address this, we propose a gated attention mechanism that dynamically assigns importance weights to each view based on the global context of all four views. This allows the model to emphasize informative features while down-weighting redundant or less relevant inputs.

Let  $v_{LCC}, v_{LMLO}, v_{RCC}, v_{RMLO} \in \mathbb{R}^{D}$ be the feature vectors extracted from each view. These are concatenated into a global context vector:

\begin{equation}
    v_{global} = \text{Concat}(v_{LCC}, v_{LMLO}, v_{RCC}, v_{RMLO})
\end{equation}

This vector is passed through a two-layer MLP with ReLU activation to produce logits, followed by Softmax to yield attention weights $\boldsymbol{\alpha} = [\alpha_{LCC}, \alpha_{LMLO}, \alpha_{RCC}, \alpha_{RMLO}]$, where each $\alpha_i \in [0, 1]$ and $\sum_{i} \alpha_i = 1$:

\begin{equation}
    \boldsymbol{\alpha} = \text{Softmax}(\text{MLP}(v_{global}))
\end{equation}

Each feature vector is then scaled by its corresponding weight and concatenated (see Figure \ref{fig:resenet}):

\begin{equation}
\begin{split}
v_{\text{fused}} = \text{Concat}(&\alpha_{LCC} \cdot v_{LCC}, \, \alpha_{LMLO} \cdot v_{LMLO}, \\
                                 &\alpha_{RCC} \cdot v_{RCC}, \, \alpha_{RMLO} \cdot v_{RMLO})
\end{split}
\end{equation}

This attention-weighted concatenation preserves the full feature space while enabling data-driven, case-specific fusion. It also improves the interpretability of the model: Attention scores $\boldsymbol{\alpha}$ offer information on the model priorities, aligned with the objectives of explainable AI in clinical practice.

\section{Dataset}

Several mammography datasets have been released in recent years, each offering different combinations of diagnostic and BI-RADS annotations. For this study, we selected datasets that satisfied four key criteria: (1) a sufficiently large cohort size, (2) availability of multi-view images with a substantial number of cases containing all four standard views, (3) inclusion of both diagnostic and BI-RADS labels, and (4) presence of normal cases (BI-RADS 1). Among widely used datasets, MIAS \cite{suckling1994mammographic}, INBreast \cite{moreira2012inbreast}, CBIS-DDSM \cite{lee2017curated}, CSAW-CC \cite{dembrower2020multi}, VinDr-Mammo \cite{nguyen2023vindr}, and CMMD \cite{cui2021chinese}, only the TOMPEI-CMMD \cite{kashiwada2025tompei} fulfilled all these criteria. This version comprises 1,775 studies spanning BI-RADS categories 1 to 5 and includes corresponding diagnostic labels (normal, benign, malignant).

The foundation of this study is the publicly available Chinese Mammography Database (CMMD) \cite{cui2021chinese}, a large-scale collection hosted on The Cancer Imaging Archive (TCIA). The original CMMD dataset consists of 3,728 breast images from 1,775 unique patients, acquired with a GE Senographe DS mammography system between 2012 and 2016. A subsequent key extension to this dataset, the TOMPEI-CMMD \cite{kashiwada2025tompei}, provided valuable corrections to image labeling and expert-annotated lesion segmentation, though it focused exclusively on the MLO views

\begin{figure}
    \centering
    \includegraphics[width=0.75\linewidth]{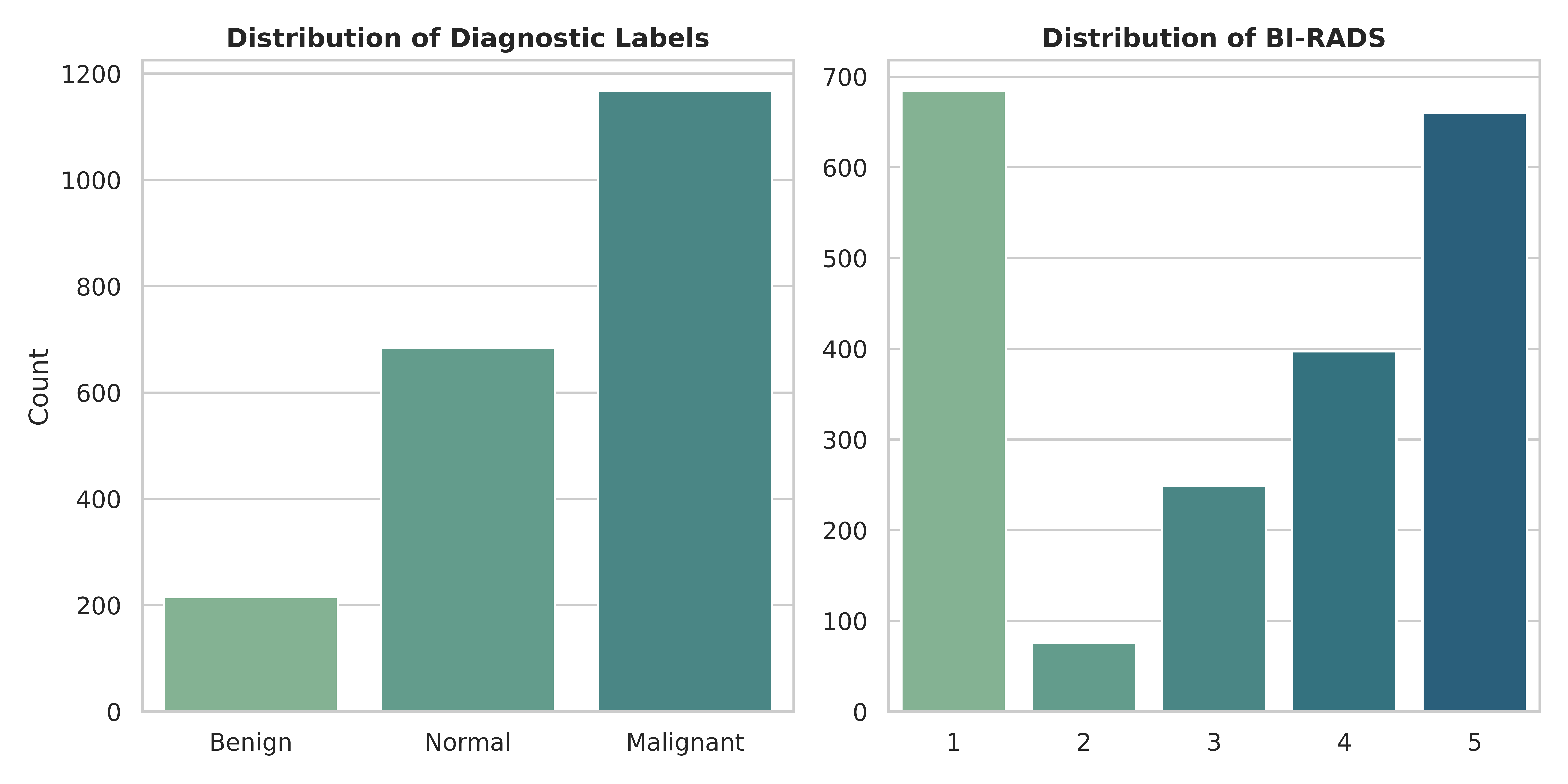}
    \caption{Breast-based diagnostic labels and BI-RADS class distribution in the TOMPEI-CMMD dataset.}
    \label{fig:distribution}
\end{figure}

To enable a clinically realistic, multi-view, multi-task framework, we made two key modifications. First, we distinguished between complete four-view cases (R-CC, R-MLO, L-CC, L-MLO) and incomplete cases (e.g., single-side only), allowing the model to reflect real-world variability while leveraging breast asymmetry. Unlike prior studies that discarded incomplete cases, we preserved them to enhance clinical applicability. Second, we jointly used BI-RADS (1–5) and diagnostic labels (normal, benign, malignant) in a unified multi-task setup, emulating the sequential decision-making process in clinical screening. To our knowledge, this is the first study applying joint BI-RADS and diagnostic classification on this dataset. After excluding cases with missing views, labels, or poor image quality, the final cohort included 1,360 patients. Data were split at the patient level into 80\% training (1,088 patients) and 20\% validation (272 patients) to avoid data leakage. The distribution of classes across breast sides is illustrated in Figure \ref{fig:distribution}.

\subsection{Preprocessing and Data Augmentation}

All images from the right breast were horizontally flipped to ensure consistent orientation across all views. Subsequently, all images were resized to a fixed resolution of $512 \times 512$ pixels. To enhance model generalization and mitigate overfitting, a series of data augmentation techniques were applied to the training set. These augmentations included random affine transformations, comprising rotations within $\pm10^\circ$, translations within $\pm5\%$, and scaling between 95\% and 105\% and random horizontal flips with a probability of 0.5.

For cases with missing views, zero tensors were used to replace the absent images, and a label of -1 was assigned for both tasks. These instances were excluded from loss computation during training and omitted from performance evaluation during testing. For the diagnostic label task, labels corresponding to normal and benign cases were grouped under a single class (label 0) to represent non-malignant cases.

\section{Experimental Setting}
This section provides an overview of the experimental setup, including the loss functions used during training, detailed model configurations, implementation specifics, and the evaluation metrics employed for performance assessment.

\subsection{Loss Function}
The training process is driven by a composite loss function, $\mathcal{L}_{\text{total}}$, which is a weighted sum of the losses from each individual task. The loss for a single task (e.g., label classification) is computed by averaging the Cross-Entropy (CE) loss from the left and right prediction heads. To address the significant class imbalance present in the dataset, the Cross-Entropy loss for each task was weighted by the inverse frequency of each class in the training set.  The total loss is formulated as:

\begin{equation}
    \mathcal{L}_{\text{total}} = w_{\text{label}} \cdot \mathcal{L}_{\text{label}} + w_{\text{birads}} \cdot \mathcal{L}_{\text{birads}}
\end{equation}
where $w_{\text{label}}$ and $w_{\text{birads}}$ are task-specific weights, set to 0.5 each to ensure equal importance during training. The individual task losses are defined as:
\begin{equation}
    \mathcal{L}_{\text{label}} = \frac{1}{2} \left( \text{CE}(y_{L, \text{label}}, \hat{y}_{L, \text{label}}) + \text{CE}(y_{R, \text{label}}, \hat{y}_{R, \text{label}}) \right)
\end{equation}
\begin{equation}
    \mathcal{L}_{\text{birads}} = \frac{1}{2} \left( \text{CE}(y_{L, \text{birads}}, \hat{y}_{L, \text{birads}}) + \text{CE}(y_{R, \text{birads}}, \hat{y}_{R, \text{birads}}) \right)
\end{equation}

\subsection{Experimental Setup}
To provide a thorough comparative analysis, we designed a series of experiments to evaluate various backbone architectures under different levels of task complexity. To determine the optimal number of VSSM blocks for each model, we performed a grid search. For the pure VSSM models, we evaluated configurations with 2 to 10 blocks and observed that the best performance was consistently achieved with four successive VSSM blocks. In the hybrid models, we incorporated half the number of VSSM layers used in the pure models to balance performance and model complexity. Since the backbone networks were initialized with pre-trained weights requiring 3-channel input, each grayscale mammography image was replicated across three channels to meet the input requirements of the pre-trained layers. 

Three subsets of the dataset were selected: (1) a high-contrast classification task including only BI-RADS categories 1 and 5; (2) a moderately ambiguous classification task comprising BI-RADS categories 1, 3, and 5; and (3) the most challenging setting, which involves the full dataset, characterized by significant class imbalance, inter-class similarities, and intra-class variability. This structured approach enables a direct comparison of model performance across different difficulty levels and allows us to assess how diagnostic ambiguity affects each architecture. All models were evaluated in both single-task and multi-task settings to provide a more comprehensive analysis of the proposed framework’s capabilities.

\subsection{Implementation Details}
All models were trained using the PyTorch deep learning framework. The training was conducted on a single NVIDIA A100 GPU. We used the AdamW optimizer  with a weight decay of 0.05. The initial learning rate was set to $1 \times 10^{-4}$ for ResNet-18 and the Hybrid model, and a lower rate of $1 \times 10^{-5}$ for the VSSM models, which were found to be more sensitive due to the lack of any pre-training for this model. An adaptive schedule for learning rate was employed, which it was reduced by a factor of 0.5 if the monitored validation metric (a combined score of label AUC and BI-RADS AUC) did not improve for 10 epochs. In addition to weight decay, we used Dropout (p=0.5) in the fusion head and gradient clipping with a maximum norm of 1.0 to ensure training stability, particularly for the VSSM models. Each model was trained for a maximum of 100 epochs, with the best performing checkpoint on the validation set saved for final analysis.

\subsection{Evaluation Metrics}

To comprehensively assess the performance of our multi-task classification model, we adopted two widely-used evaluation metrics: the Area Under the Receiver Operating Characteristic Curve (AUC) and the Macro-averaged F1 score (F1-macro). These metrics were selected based on the nature of each classification task and their ability to provide a robust evaluation under class imbalance, which is prevalent in the utilized dataset. The AUC quantifies the model's ability to discriminate between classes by measuring the area under the Receiver Operating Characteristic (ROC) curve. It is applicable to both binary and multiclass classification tasks. AUC measures the model’s ability to rank positive instances higher than negative ones across all possible thresholds. It is threshold-independent and focuses on separation of classes. To complement AUC, we also report the macro-averaged F1 score (F1-macro), which captures the balance between precision and recall for each class independently and then averages them. It is threshold-dependent and sensitive to class imbalance.

\section{Results and Discussion}
This section presents a comprehensive analysis of the experimental outcomes This section presents a comprehensive analysis of the experimental results obtained from our multi-task learning framework. We begin by reporting the quantitative performance of five distinct models across three experimental settings, each reflecting increasing levels of clinical complexity, under both single-task and multi-task training paradigms. This is followed by an in-depth discussion that synthesizes the findings, highlighting key architectural trade-offs and exploring their potential clinical implications. To evaluate the learning capabilities and robustness of each backbone, we designed three primary experimental settings. The diagnostic "Label Task" (Normal vs. Malignant) was kept consistent across all runs, while the complexity of the BI-RADS classification task was progressively increased.

\textbf{High-Contrast Binary Classification:} This initial experiment tests the models' ability to distinguish between clearly normal (BI-RADS 1) and highly suspicious for malignancy (BI-RADS 5) cases. The results are summarized in Table \ref{tab:binary_groups}. In the high-contrast scenario, the ResNet-50 model outperformed ResNet-18 in the multi-task setting; however, its performance was slightly lower in single-task training, highlighting the advantage of multi-task learning in leveraging complementary information and promoting more generalizable feature representations. A similar trend was observed with the VSSM models: the view-specific variant was dominant in multi-task learning, while the shared version performed better under single-task training. Notably, view-specific models have nearly twice the number of parameters compared to their shared counterparts, making them more prone to overfitting, an issue that multi-task learning can mitigate by acting as an implicit regularizer. The proposed hybrid model outperformed all other architectures across all metrics, achieving an AUC of 0.9967 and an F1 score of 0.9830 in multi-task learning, results nearly identical to those in single-task training. Among hybrid models, the shared variant surpassed the view-specific version, and in both cases, multi-task learning yielded better performance than single-task learning in most evaluations.

\begin{table*}[ht!]
    \centering
    \caption{Best Validation Performance on Binary Tasks (Normal vs. Malignant \& BI-RADS 1 vs. 5). Best performance in each column is in bold.}
    \label{tab:binary_groups}
    \renewcommand{\arraystretch}{1.2}
    \resizebox{\textwidth}{!}{%
    \begin{tabular}{@{}lcccccccc@{}}
        \toprule
        \multirow{3}{*}{\textbf{Model Backbone }} 
        & \multicolumn{4}{c}{\textbf{Multi-task}} 
        & \multicolumn{4}{c}{\textbf{Single-task}} \\
        \cmidrule(lr){2-5} \cmidrule(lr){6-9}
        & \textbf{AUC} & \textbf{F1-Macro} & \textbf{AUC} & \textbf{F1-Macro} 
        & \textbf{AUC} & \textbf{F1-Macro} & \textbf{AUC} & \textbf{F1-Macro} \\
        & \multicolumn{2}{c}{\textbf{\{Normal, Malignant\}}} & \multicolumn{2}{c}{\textbf{BI-RADS: \{1, 5\}}} 
        & \multicolumn{2}{c}{\textbf{\{Normal, Malignant\}}} & \multicolumn{2}{c}{\textbf{BI-RADS: \{1, 5\}}} \\ 
        \midrule
        ResNet-18  & 0.9261 & 0.8447 & 0.9555 & 0.9383 & 0.9289 & 0.8675 & 0.9580 & 0.9353 \\ 
        ResNet-50 & 0.9304 & 0.8727 & 0.9578 & 0.9382 & 0.9245 & 0.8872 & 0.9571 & 0.9300 \\ \midrule
        VSSM (\texttt{Shared}) & 0.9319 & 0.8731 & 0.9775 & 0.9208 & 0.9439 & 0.8933 & 0.9877 & 0.9390 \\
        VSSM (\texttt{View-specific}) & 0.9350 & 0.8785 & 0.9820 & 0.9294 & 0.9188 & 0.8521 & 0.9706 & 0.9000 \\ \midrule
        \textbf{Hybrid ResNet-VSSM (\texttt{Shared})} & \textbf{0.9686} & \textbf{0.9259} & \textbf{0.9967} & \textbf{0.9830} & 0.9533 & \textbf{0.9096} & \textbf{0.9982} & \textbf{0.9829} \\
        \textbf{Hybrid ResNet-VSSM (\texttt{View-specific})} & 0.9566 & 0.8978 & 0.9958 & 0.9717 & \textbf{0.9577} & 0.8966 & 0.9939 & 0.9525 \\
        \bottomrule
    \end{tabular} }
\end{table*}

\textbf{Introducing Ambiguity:} This experiment increases clinical complexity by including the intermediate BI-RADS 3 cases, testing the models' ability to handle diagnostic uncertainty. The results for this experiment are summarized in Table \ref{tab:ternary_groups}. Similar to the high-contrast scenario, the proposed hybrid model consistently achieves the highest scores across all tasks and metrics. Notably, the shared hybrid variant attains the best performance in BI-RADS classification under the multi-task setting, with an AUC of 0.9781 and an F1 score of 0.7790, demonstrating the advantage of combining local feature extraction with global context modeling. ResNet-18 and ResNet-50 exhibit nearly similar performance, with their largest discrepancy observed in the binary-task F1 score, which is notably mitigated in the multi-task setup. VSSM models generally underperform compared to their hybrid counterparts, particularly in the BI-RADS classification tasks. For example, in the multi-task setting, the shared VSSM achieves an F1 score of 0.7328, while the shared hybrid reaches 0.7790. Across most backbones, multi-task learning improves both AUC and F1 scores for the ternary BI-RADS classification relative to single-task learning. This trend underscores the benefit of auxiliary supervision in helping the model develop more robust and transferable feature representations, particularly valuable for capturing the subtle distinctions inherent in BI-RADS categorization.

\begin{table*}[ht!]
    \centering
    \caption{Best Validation Performance on Ternary BI-RADS Task (1 vs. 3 vs. 5). Best performance in each column is in bold.}
    \label{tab:ternary_groups}
    \renewcommand{\arraystretch}{1.2}
    \resizebox{\textwidth}{!}{%
    \begin{tabular}{@{}lcccccccc@{}}
        \toprule
        \multirow{3}{*}{\textbf{Model Backbone}} 
        & \multicolumn{4}{c}{\textbf{Multi-task}} 
        & \multicolumn{4}{c}{\textbf{Single-task}} \\
        \cmidrule(lr){2-5} \cmidrule(lr){6-9}
        & \textbf{AUC} & \textbf{F1-Macro} & \textbf{AUC} & \textbf{F1-Macro} 
        & \textbf{AUC} & \textbf{F1-Macro} & \textbf{AUC} & \textbf{F1-Macro} \\
        & \multicolumn{2}{c}{\textbf{\{Normal, Malignant\}}} & \multicolumn{2}{c}{\textbf{BI-RADS: \{1, 3, 5\}}} 
        & \multicolumn{2}{c}{\textbf{\{Normal, Malignant\}}} & \multicolumn{2}{c}{\textbf{BI-RADS: \{1, 3, 5\}}} \\ 
        \midrule
        ResNet-18  & 0.9302 & 0.8731 & 0.9414 & 0.7762 & 0.9247 &  0.8489 & 0.9378 & 0.7588 \\ 
        ResNet-50 & 0.9307 & 0.8784 & 0.9411 & 0.7623 & 0.9235& 0.8747 & 0.9353 & 0.7320 \\ \midrule
        VSSM (\texttt{Shared}) & 0.9362 & 0.8818 & 0.9614 & 0.7328 & 0.9275 & 0.8667 & 0.9483 & 0.6985 \\
        VSSM (\texttt{View-specific}) & 0.926 & 0.8618 & 0.9584 & 0.7388 & 0.9272 & 0.8600 & 0.89505 & 0.7055\\ \midrule
        \textbf{Hybrid ResNet-VSSM (\texttt{Shared})} & \textbf{0.9629} & 0.8858 & \textbf{0.9781} & \textbf{0.7790} & \textbf{0.9616} & 0.8923 & \textbf{0.9761} & 0.7552 \\
        \textbf{Hybrid ResNet-VSSM (\texttt{View-specific})} & 0.9529 & \textbf{0.8907} & 0.9708 & 0.7332 & 0.9556 & \textbf{0.9005} & 0.9742 & \textbf{0.7729} \\ 
        \bottomrule
    \end{tabular} }
\end{table*}

\textbf{Full Multi-Class Challenge (All BI-RADS Classes):}
This final experiment represents the most challenging and clinically realistic scenario, in which models are tasked with classifying the full categories of BI-RADS scores (1–5). This setting is characterized by high inter-class similarity, substantial intra-class variability, and significant class imbalance, factors that collectively make both training and decision-making particularly difficult for deep learning models. The results for this experiment are summarized in Table \ref{tab:all_groups}. Consistent with previous classification tasks, the hybrid ResNet-VSSM models outperform all baselines across both classification tasks. The shared variant of the hybrid model achieves the highest multi-task performance for the five-class BI-RADS classification, with an AUC of 0.9249 and an F1 score of 0.4597, which further improves to 0.9265 AUC and 0.4904 F1 in the single-task setting. While binary classification yields relatively high AUC scores (up to 0.9607), F1 scores for the five-class BI-RADS task remain considerably lower across all models—typically below 0.5. This performance gap underscores the inherent challenge of distinguishing between intermediate BI-RADS categories, which often exhibit subtle and ambiguous radiological characteristics.

\begin{table*}[ht!]
    \centering
    \caption{Best Validation Performance on Ternary BI-RADS Task (All classes). Best performance in each column is in bold.}
    \label{tab:all_groups}
    \renewcommand{\arraystretch}{1.2}
    \resizebox{\textwidth}{!}{%
    \begin{tabular}{@{}lcccccccc@{}}
        \toprule
        \multirow{3}{*}{\textbf{Model Backbone}} 
        & \multicolumn{4}{c}{\textbf{Multi-task}} 
        & \multicolumn{4}{c}{\textbf{Single-task}} \\
        \cmidrule(lr){2-5} \cmidrule(lr){6-9}
        & \textbf{AUC} & \textbf{F1-Macro} & \textbf{AUC} & \textbf{F1-Macro} 
        & \textbf{AUC} & \textbf{F1-Macro} & \textbf{AUC} & \textbf{F1-Macro} \\
        & \multicolumn{2}{c}{\textbf{\{Normal, Malignant\}}} & \multicolumn{2}{c}{\textbf{BI-RADS: \{1, 2, 3, 4, 5\}}} 
        & \multicolumn{2}{c}{\textbf{\{Normal, Malignant\}}} & \multicolumn{2}{c}{\textbf{BI-RADS: \{1, 2, 3, 4, 5\}}} \\ 
        \midrule
        ResNet-18  & 0.9238 & 0.8767 & 0.8872 & 0.4464 & 0.9274 &  0.8746 & 0.8857 & 0.4740 \\ 
        ResNet-50 & 0.9254 & 0.8604 & 0.8909 & 0.4382 & 0.9234 & 0.8768 & 0.8925 & 0.4557 \\ \midrule
        VSSM (\texttt{Shared}) & 0.9240 & 0.8529 & 0.8845 &  0.4204 & 0.9262  & 0.8855 & 0.8860 & 0.3948 \\
        VSSM (\texttt{View-specific}) & 0.9297 & 0.8617 & 0.8880 & 0.3832 & 0.0.9215 & 0.8803 & 0.8766 & 0.3713 \\ \midrule
        \textbf{Hybrid ResNet-VSSM (\texttt{Shared})} & \textbf{0.9607} & \textbf{0.9261} & \textbf{0.9249} & \textbf{0.4597} & \textbf{0.9586} & 0.8954 & \textbf{0.9265} & \textbf{0.4904} \\
        \textbf{Hybrid ResNet-VSSM (\texttt{View-specific})} & 0.9571 & 0.8944 & 0.9190 & 0.4109 & 0.9555 & \textbf{0.8984} & 0.9059 & 0.4556\\ 
        \bottomrule
    \end{tabular} }
\end{table*}

Unlike the clearer improvements observed in the ternary BI-RADS setting, the benefits of multi-task learning in the five-class scenario are less consistent. This may suggest that auxiliary binary tasks offer limited benefit when the target classification involves fine-grained distinctions or when class imbalance becomes more severe. Nevertheless, shared versions of both VSSM and hybrid architectures outperform their view-specific counterparts in this setting. For example, the shared hybrid model improves BI-RADS classification AUC from 0.9190 to 0.9249 and F1 score from 0.4109 to 0.4597 in multi-task learning. These findings suggest that the proposed VSSM framework, when implemented in a shared configuration, is capable of learning robust representations despite anatomical variation across views. Moreover, shared models benefit from a larger effective training sample size and demonstrate reduced susceptibility to overfitting compared to view-specific models.

\section{Conclusion }

In this work, we introduced a novel multi-view, multi-task deep learning framework for mammography image analysis that combines the strengths of CNNs and VSSMs. Our hybrid CNN-VSSM architecture effectively captures both local spatial details and long-range global dependencies, addressing the limitations of conventional CNNs and the high computational complexity of transformer-based methods. Furthermore, we designed a gated attention-based fusion module that dynamically integrates multi-view information, offering robustness in scenarios with incomplete data and enhancing the interpretability of the model’s predictions.

We conducted comprehensive experiments across progressively complex diagnostic scenarios, ranging from binary classification to full five-class BI-RADS categorization. The results demonstrated that our proposed hybrid models, particularly those employing shared representations, consistently achieved superior performance in both AUC and F1-Macro scores. This confirms the effectiveness of integrating local CNN-based features with global contextual modeling via VSSM. Binary classification tasks achieved the highest performance, with shared hybrid models approaching near-perfect scores (AUC > 0.99 and F1-Macro > 0.98), indicating that these coarse-grained distinctions are well captured by the models. However, as the classification complexity increased, performance declined accordingly. In the ternary BI-RADS task, although hybrid models remained dominant, F1 scores dropped below 0.78. The five-class BI-RADS classification proved most challenging, with even the best models yielding F1 scores below 0.5. These results highlight the inherent difficulty of modeling subtle radiological variations and the intra-class ambiguity present in intermediate BI-RADS categories.

Multi-task learning generally provided clear benefits in binary and ternary classification settings, improving both robustness and generalization. However, in the five-class task, these advantages were less consistent, with single-task models occasionally outperforming their multi-task counterparts. Additionally, while view-specific models occasionally yielded strong results in simpler binary tasks, shared models demonstrated superior robustness and generalizability in more complex classification scenarios.

In summary, our work highlights the benefits of hybrid and multi-task architectures in medical imaging and paves the way for future research in more clinically realistic and scalable AI systems. By combining efficient feature encoding, global contextual reasoning, and intelligent multi-view fusion, the proposed framework represents a promising step toward reliable and interpretable AI-driven decision support tools in breast cancer screening and diagnosis.

\section*{Acknowledgments}
This work is supported under the International Research Collaboration Co-Fund (IRCC) between Qatar University and University of Hyogo.   Grant number IRCC-2025-633.

\bibliographystyle{ieeetr}
\bibliography{refs.bib}

\end{document}